\documentclass[aps,prl,twocolumn,showpacs,superscriptaddress,groupedaddress]{revtex4}

\usepackage{graphicx,amsmath,amssymb}
\usepackage{color}

\begin{document}
\title{Replica symmetry breaking in trajectories of a driven Brownian particle}
\author{Masahiko Ueda}
\email[]{ueda@ton.scphys.kyoto-u.ac.jp}
\address{Department of Physics, Kyoto University, Kyoto 606-8502, Japan}
\author{Shin-ichi Sasa}
\email[]{sasa@scphys.kyoto-u.ac.jp}
\address{Department of Physics, Kyoto University, Kyoto 606-8502, Japan}

\date{\today}

\begin{abstract}
We study a Brownian particle passively driven by a field obeying the noisy Burgers equation. 
We demonstrate that the system exhibits replica symmetry breaking in the path ensemble with the initial position of the particle being fixed. 
The key step of the proof is that the path ensemble with a modified boundary condition can be exactly mapped to the canonical ensemble of directed polymers.
\end{abstract}

\pacs{05.40.-a, 05.70.Fh, 64.70.P-}
\maketitle


{\em Introduction.---}
Fluctuation and transportation in active environments have 
gradually attracted attention in recent studies of non-equilibrium physics
and biophysics \cite{SaxJac1997, WuLib2000, Eweet2005, Buret2005, MTSM2007, LiuGor2008, WABG2009, Guoet2014}. 
Among many problems, 
the prediction of the transportation properties of biological materials in 
living cells is a challenging problem in theoretical physics,
and the possibility of such prediction is a significant step in the control of cells \cite{SaxJac1997, Eweet2005, Buret2005, WABG2009, Guoet2014}. 
In order to solve the problem, recalling the theory of Brownian motion in equilibrium environments \cite{Ein1956}, we seek for a universal principle for fluctuation and transportation in active environments by the analysis of trajectories of tracer particles.
In particular, currently, it is significant to discover a key concept that is helpful in studying apparently complicated phenomena. 
We thus study the motion of a tracer particle in a simple system.


When we focus on the trajectories of a tracer particle, the 
most fundamental quantity is the displacement of the particle.
For example, the mean-squared displacement as a function of time
classifies diffusion behaviors such as superdiffusion
\cite{BGKPR1990, BasIom2004, LiuGor2008} and 
subdiffusion \cite{WEKN2004, Wonet2004}. 
More generally, the statistical properties of displacement during a finite time interval
are characterized by the cumulant generating function of the displacement. 
This is often referred to as the {\it dynamical free energy}  
of displacement, which is associated with a statistical mechanical 
framework of trajectories \cite{BecSch1993, MGC2005}. 
Such a framework can 
be developed for other quantities such as activity \cite{GJLPDW2007, GJLPDW2009, HJGC2009, JacGar2010}
and Lyapunov exponent \cite{GKLT2011, LLKT2013, LSTW2015}. 
In this context, as a remarkable result, the study of the dynamical free energy of activity in glassy systems led to the discovery of the first-order transition,
wherein dynamical heterogeneity is described as the coexistence
of active and inactive phases in space-time at the first-order 
transition point. 
Since analysis of the dynamical free energy 
can reveal the singular nature of trajectories, it is useful 
to discover new phenomena associated with hidden singularities 
of the trajectories of a tracer particle in active environments.


In this Letter, we study the dynamical free energy of {\it overlap}.
We here briefly review the concept of overlap in spin glass theory \cite{MPV1987}. 
Suppose that two independent and identical mean-field spin-glass systems (replicas) are prepared. 
Then, the distribution function of the overlap, which represents the similarity between the two spin configurations, takes a non-trivial form in the spin glass phase, reflecting the existence of several stable configurations.
This corresponds to {\it replica symmetry breaking} (RSB). 
The concept of RSB has so far been applied to information theory and computational complexity theory \cite{MezMon2009}. 
Here, we apply this method of RSB detection to the trajectories of two tracer particles \cite{Mik}. 
That is, we prepare two independent particles and define the overlap as the similarity between the trajectories of both particles. 
We identify RSB by the existence of a non-trivial feature of the distribution function of the overlap. 
Therefore, the dynamical phase transition reported in this Letter is characterized by RSB in the path ensemble.


Concretely, we study an active environment 
obeying the noisy Burgers equation \cite{BecKha2007}.
The noisy Burgers equation was introduced as a toy model 
of turbulence, and it is equivalent to the Kardar--Parisi--Zhang (KPZ) 
equation \cite{KPZ1986}, which has been extensively studied \cite{FNS1977, MHKZ1989, 
KMH1992, BMP1995, BerGia1997, CCDW2010, SasSpo2010, CalLed2011, HalZha1995, Mauet1997, TSSS2011}.
In this study, we find that the overlap between the trajectories 
of two tracer particles obeys a non-trivial distribution, 
and we subsequently provide evidence to support the claim that 
this model exhibits RSB in the path ensemble.


{\em Model.---}
Let $x(t)$ be the one-dimensional position of a Brownian particle at time $t\in[0,\tau]$. 
The particle is assumed to be passively driven by a velocity field $u(x,t)$. 
The motion of the particle is written as 
\begin{eqnarray}
\dot{x}(t) &=& u( x(t), t ) + \xi(t),
\label{eq:passive} 
\end{eqnarray}
where $\dot{x}(t)\equiv dx(t)/dt$ and $\xi(t)$ represents the thermal noise satisfying $\left\langle \xi(t) \xi(t') \right\rangle = 2D\delta(t-t')$
with the diffusion constant $D$ for the free particle. 
The velocity field is assumed to obey the noisy Burgers equation.
By setting $u=-\partial \phi/\partial x$, the equation for $\phi$ is expressed as the KPZ equation:
\begin{eqnarray}
\frac{\partial \phi}{\partial t} &=& \nu \frac{\partial^2 \phi}{\partial x^2} 
+\frac{1}{2} \left( \frac{\partial \phi}{\partial x} \right)^2 + v(x,t),
\label{eq:KPZ}
\end{eqnarray}
where $v(x,t)$ represents zero-mean Gaussian white noise with variance $\left\langle v(x,t) v(x',t') \right\rangle  = 2B\delta(x-x')\delta(t-t')$.
More precisely, since the space-time Gaussian white noise is not properly defined, it is necessary to introduce a cutoff length $\Delta x$ in the space coordinate \cite{HalZha1995, Zha1990}. 
The derivatives in $x$ that appear in (\ref{eq:KPZ}) are interpreted as simple differences. 
Accordingly, the field $u$ acting on the particle is evaluated by linear interpolation of $\phi$. 
Hereafter, we assume that all quantities are made dimensionless. 
We set $\Delta x=0.5$, and we find later that this choice of $\Delta x$ does not lead to any problems because all other length scales in this study are larger than $\Delta x$.
For completeness, we also note the range of the particle position and the domain of the KPZ field. 
They are defined in a finite region with length $L$ under periodic boundary conditions, and $L$ is chosen to be sufficiently large such that the particle does not reach the boundaries within the observation time.


Before presenting our results, we briefly review previous studies involving this model. 
This model has been examined as an example of scalar turbulence \cite{ShrSig2000}, and it exhibits the remarkable phenomenon that non-interacting particles passively driven by the field cluster with time, wherein coalescence of valleys of the KPZ surface over time plays a key role \cite{Chi2002}. 
Theoretically, renormalization group analysis has been performed for this model \cite{DroKar2002}. 
Furthermore, clusterization in the steady state has been detected by studying density-density correlation functions \cite{NBM2005, NMB2006}. 
These previous studies indicate that particles in this model show a tendency towards localization. 
However, to the best of our knowledge, RSB in the path ensemble has never been reported. 


{\em Numerical Result.---}
We first present the numerical simulation results of the model. 
In particular, we focus on the $D=\nu$ case, which corresponds to a fixed point of the renormalization group \cite{DroKar2002, ErtKar1993}. 
Time integration is performed by employing the simplest discretization method with $\Delta t=0.01$. 
We fix the initial position of the particle as $x(0) = L/2$ with $L=10000$. 
The initial value of the field, $\phi(x,0)$, is sampled from the stationary probability distribution of (\ref{eq:KPZ}) \cite{HalZha1995}.
Numerically, we obtain the distribution as the result of a sufficiently long time evolution of $\phi$.
The parameter values are fixed as $D=\nu=1.0$ and $B=2.5$.
Here, it should be noted that there are two sources of randomness, $\xi(t)$ and $v(x,t)$. 
In calculating an ensemble average of physical quantities, we first take the statistical average over $N_1$ histories of noise $\xi$ for one realization of $u(x,t)$, and subsequently consider the configurational average over $N_2$ realizations of $v$. 
The convergence of quantities with respect to the choice of $(N_1, N_2)$ is carefully checked, and the numerical data presented below are obtained for $N_1=80000$ and $N_2=1000$. 


\begin{figure}[t]
\includegraphics[clip, width=4.25cm]{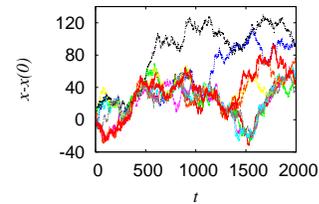}
\caption{(color online). Trajectories of a Brownian particle for one realization of $u(x,t)$. 
Ten samples are displayed.}
\label{fig:trajectories_fkpz}
\end{figure}



We start with the observation of the trajectories for one realization of $u(x,t)$. 
As displayed in Fig. \ref{fig:trajectories_fkpz},
each has the features of a typical trajectory of Brownian motion, while there exists a region in which several trajectories overlap.
We next proceed to quantify this observation. 
In the rest of this paper, the ensemble average with respect to both $\xi(t)$ and $v(x,t)$ is taken for all physical quantities.

\begin{figure}[t]
\includegraphics[clip, width=4.25cm]{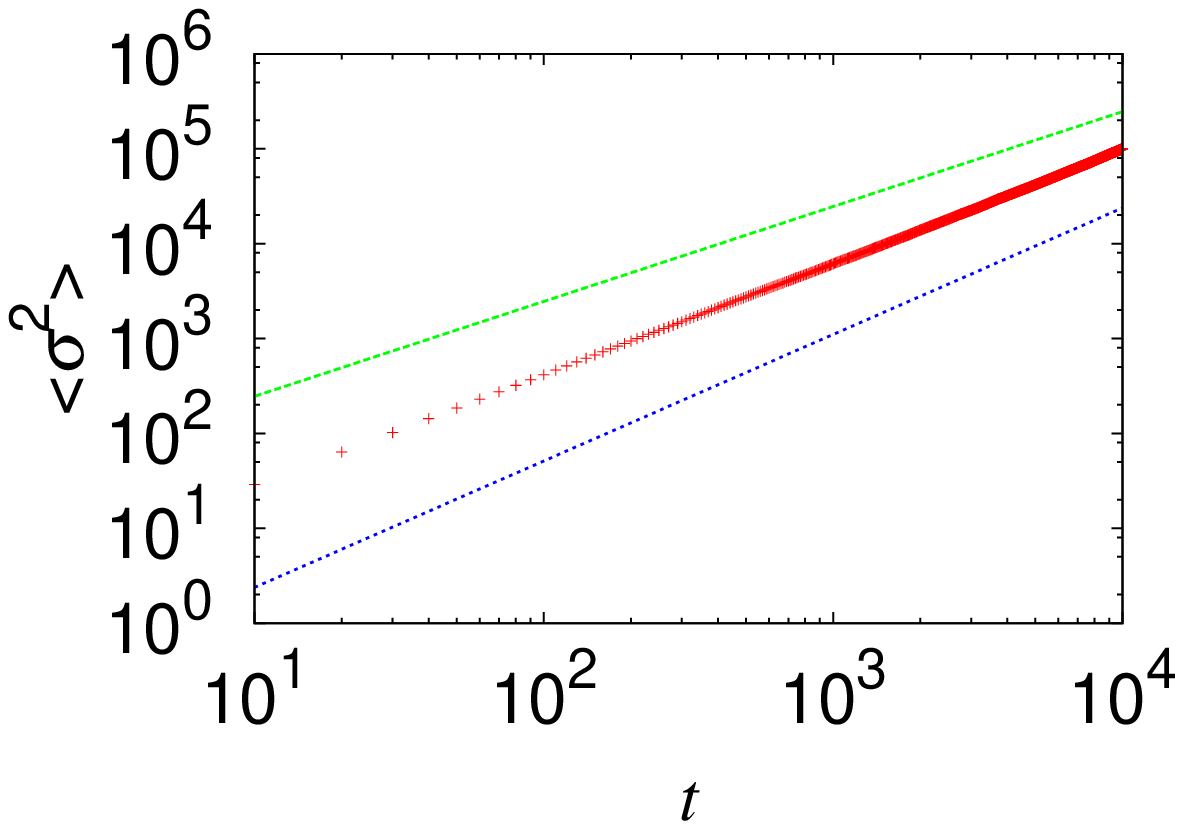}
\includegraphics[clip, width=4.25cm]{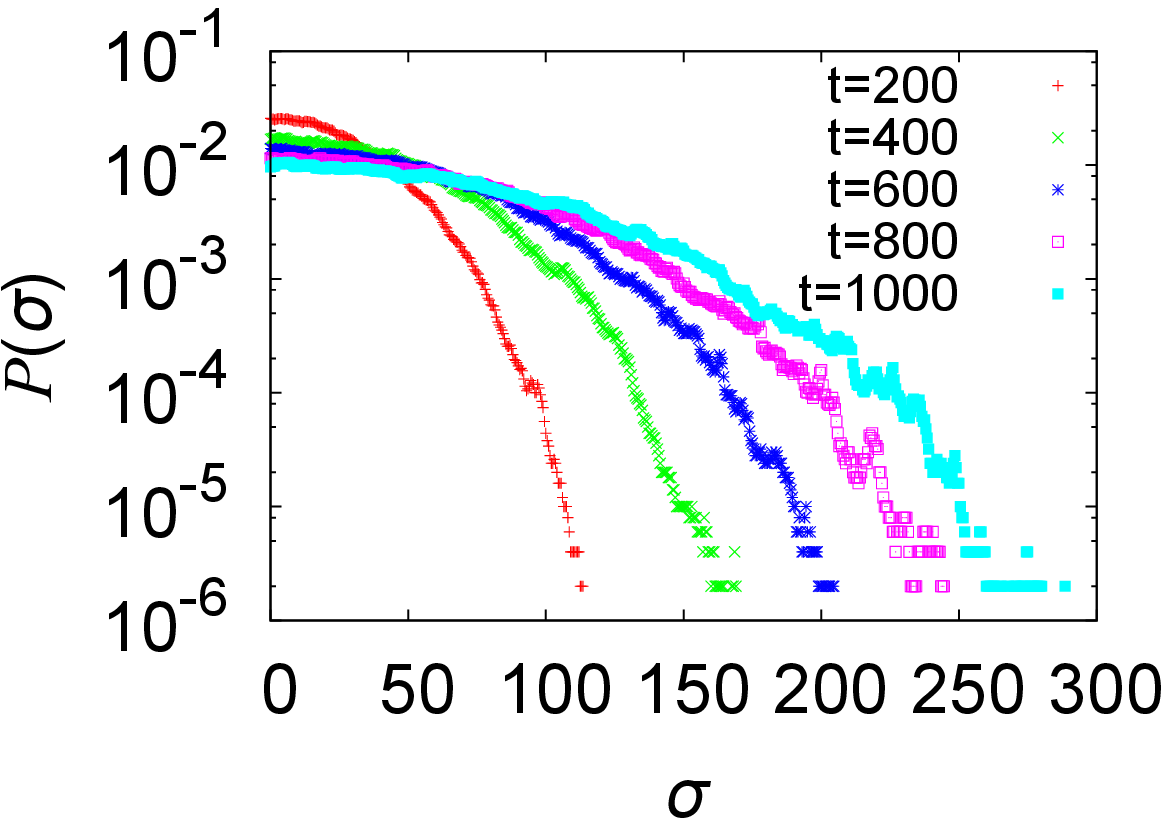}
\caption{(color online). The left figure shows the mean-squared displacement $\left\langle \sigma^2 \right\rangle$ as a function of $t$ in a log--log plot. 
The straight lines represent the normal diffusion behavior $\left\langle \sigma^2 \right\rangle \sim t$ and the anomalous diffusion behavior $\left\langle \sigma^2 \right\rangle \sim t^{4/3}$, respectively.
The right figure shows the distribution of $\sigma$ for various values of $t$. 
A logarithmic scale is used for the vertical axis.}
\label{fig:test}
\end{figure}



First, on the left side of Fig. \ref{fig:test}, the mean-squared displacement of $\sigma \equiv \left|x(t)-x(0)\right|$ as a function of $t$ is displayed. 
The particle behavior deviates from the normal diffusion type, and the anomalous diffusion $\left\langle \sigma^2 \right\rangle \sim t^{4/3}$ is observed in the long time regime, as discussed in Refs. \cite{Chi2002, DroKar2002}.
A distribution of $\sigma$ is also shown on the right side of Fig. \ref{fig:test}. 
Throughout this Letter, we use the same symbols (and colors) for such distribution functions with the same $t$.
Since the width of $P(\sigma)$ increases with time $t$, localization phenomena are not detected by studying the one-particle behavior.


\begin{figure}[t]
\includegraphics[clip, width=4.25cm]{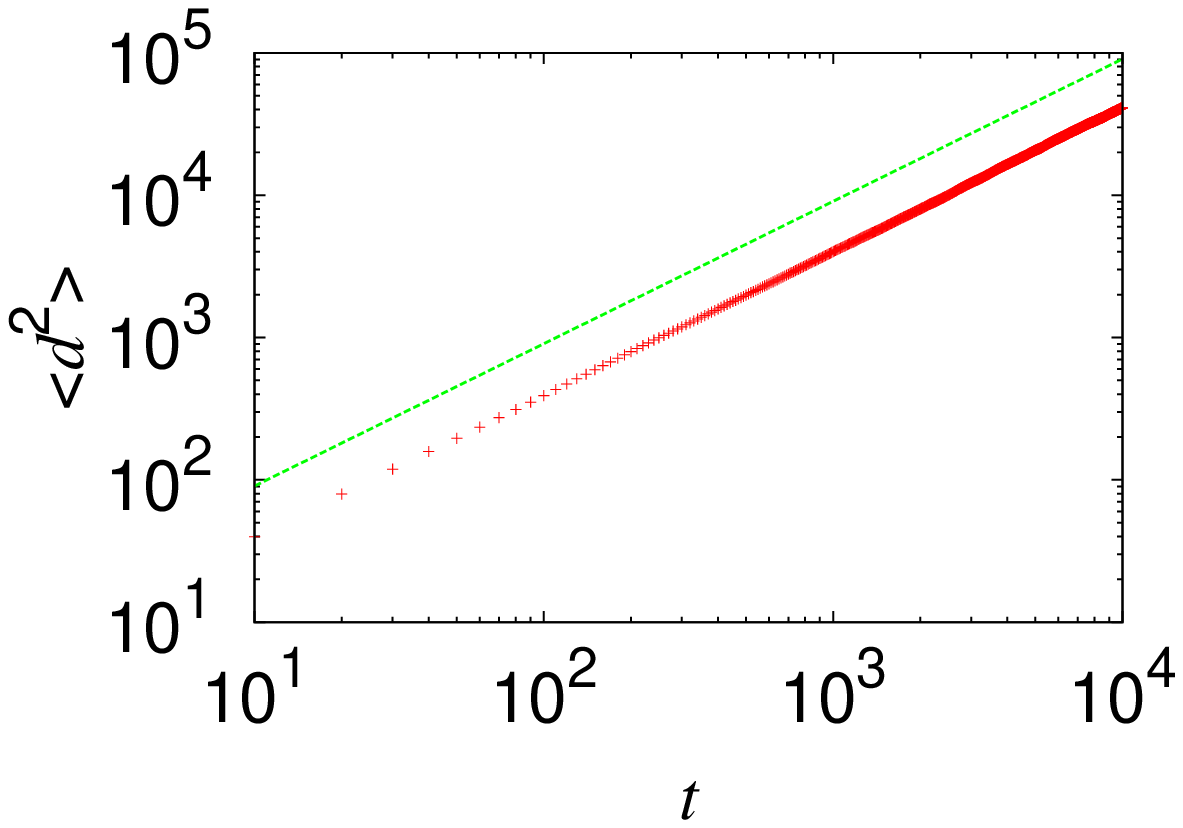}
\includegraphics[clip, width=4.25cm]{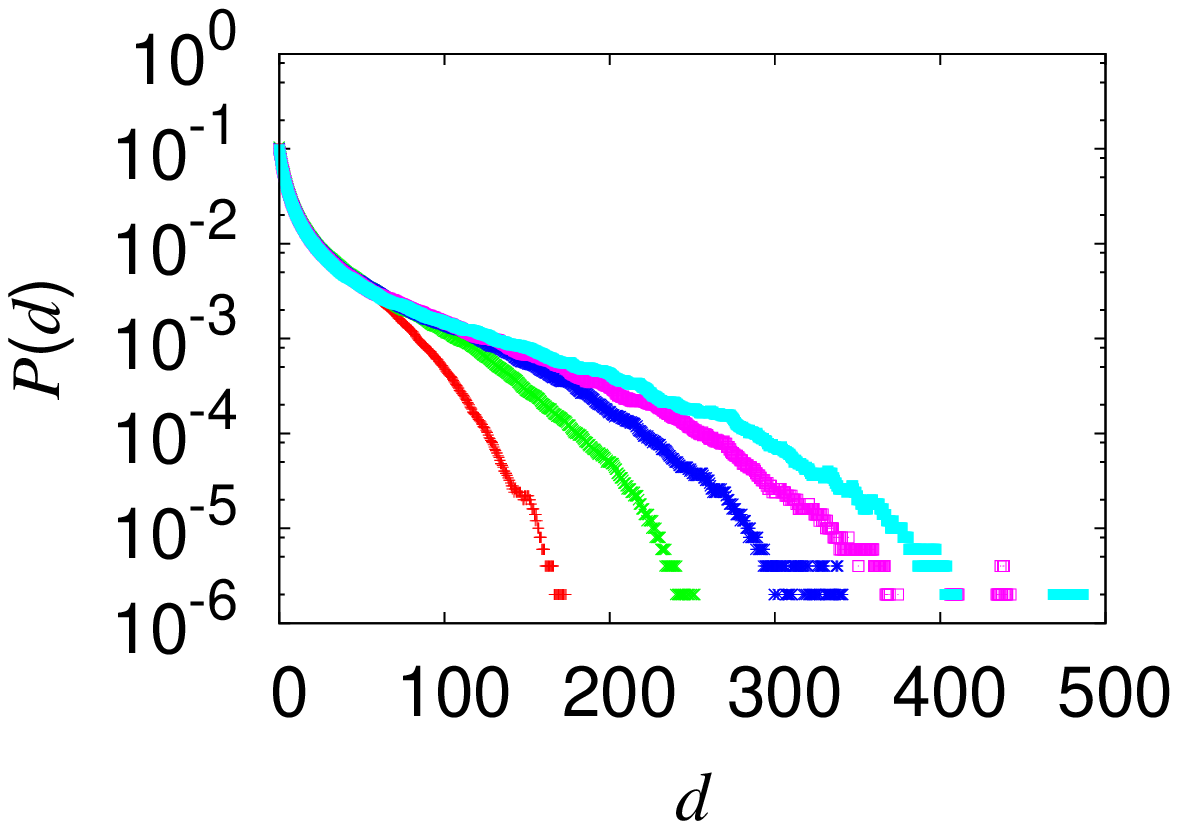}
\caption{(color online). The left figure shows the mean-squared relative distance $\left\langle d^2 \right\rangle$ as a function of $t$ in a log--log plot.
The straight line represents the normal diffusion behavior $\left\langle d^2 \right\rangle \sim t$. 
The right figure shows the distribution of $d$ for various values of $t$. 
A logarithmic scale is used for the vertical axis.}
\label{fig:test2}
\end{figure}



Next, the mean-squared value of the relative distance $d\equiv\left| x^{(1)}-x^{(2)} \right|$ for two trajectories $x^{(1)}$ and $x^{(2)}$ under the same $u(x,t)$ is investigated. 
The normal relative diffusion behavior $\left\langle d^2\right\rangle \sim t$ is clearly observed on the left side of Fig. \ref{fig:test2}. 
On the other hand, as shown on the right side of Fig. \ref{fig:test2}, the distribution of $d$ at time $t$, which is denoted by $P_t(d)$, is far from the Gaussian distribution that is obtained for free Brownian particles. 
As a characteristic feature of $P_t(d)$, it is observed that the value of $P_t(d)$ for $d < 60$ does not change over time when $t > 200$. 
The existence of such a time-independent behavior indicates that the inter-distance of two particles is not significantly larger than a certain characteristic length. 
This indication provides one piece of quantitative evidence for a localization phenomenon. 
In our study, we estimated the characteristic length scale in this region as $d_0=4.2$ \cite{SM-d0}. 
We remark here that such a diffusion scaling with a non-Gaussian distribution has attracted considerable research attention recently \cite{WABG2009, WKBG2012}.

\begin{figure}[t]
\includegraphics[clip, width=4.25cm]{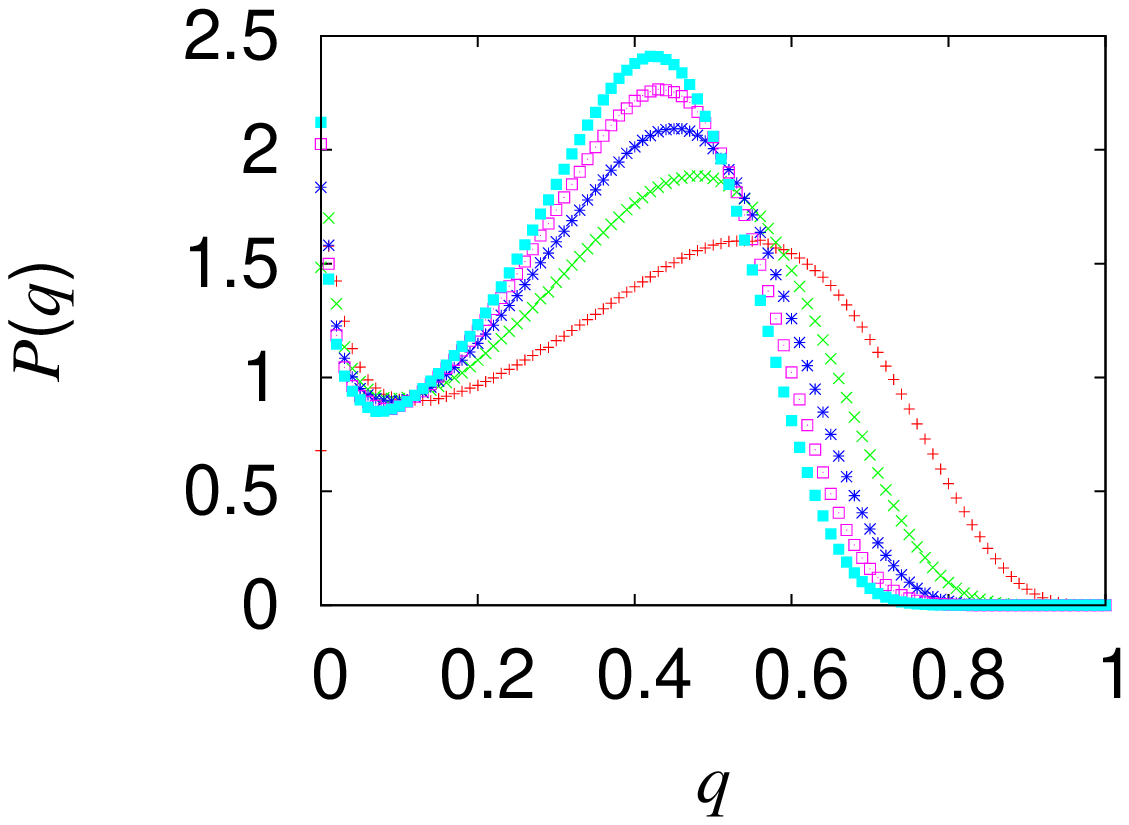}
\includegraphics[clip, width=4.25cm]{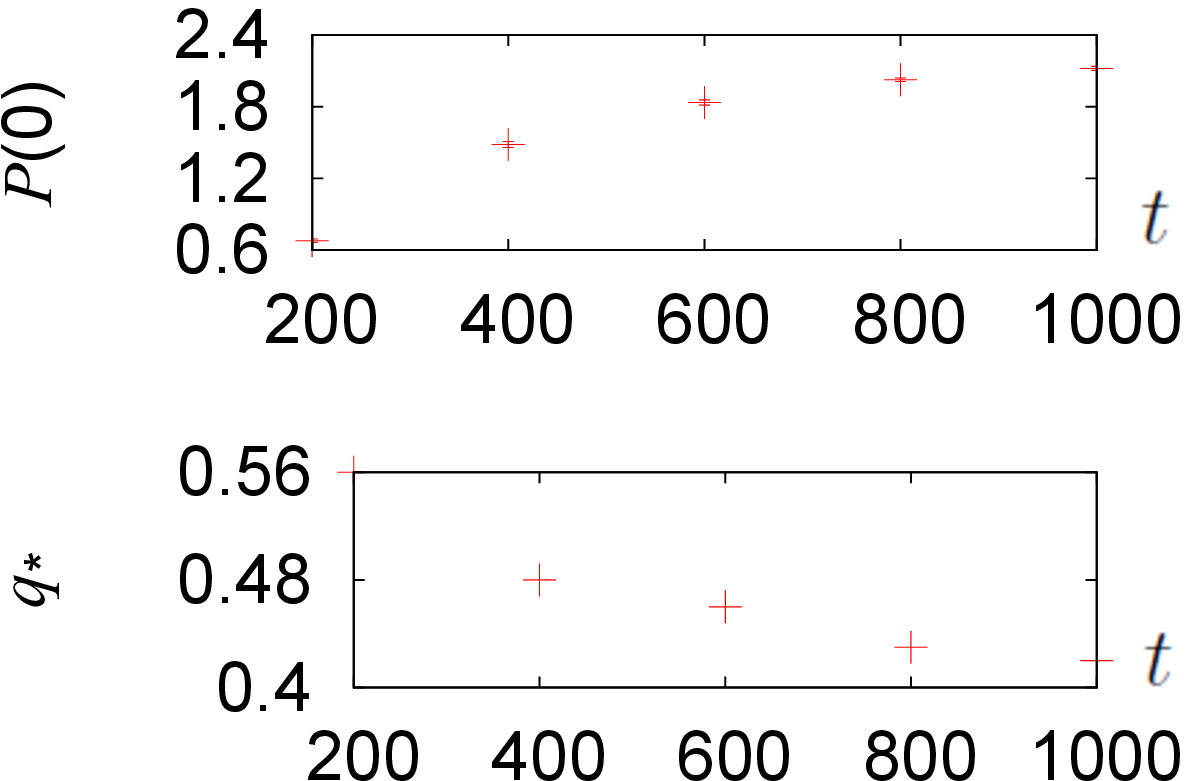}
\caption{(color online). Distribution of the overlap $q$ for various $t$ (left) and their peak values at $q=0$ and peak positions $q_*(t)$ (right).}
\label{fig:test3}
\end{figure}

In order to characterize this phenomenon as a dynamical singularity in the path ensemble, we next use an analogy with the spin glass theory.
We consider the overlap defined by
\begin{eqnarray}
q\equiv \frac{1}{M} 
\sum_{j=1}^M 
\theta\left(\ell-\left|x^{(1)}(j\Delta t)-x^{(2)}(j\Delta t) \right|\right),
\label{eq:overlap}
\end{eqnarray}
for two trajectories $x^{(1)}$ and $x^{(2)}$ under the same $u(x,t)$.
Here, $M\equiv t/\Delta t$ and we have introduced the length scale $\ell$ characterizing the localization. 
From the above discussion, $\ell$ should be close to $d_0$. 
For simplicity, we set $\ell=5$.  
On the left side of Fig. \ref{fig:test3}, we show the distribution function of the overlap, $P(q)$, for different values of $t$. 
There are two peaks at $q=0$ and $q=q_*(t)$ for every $t$, which is a characteristic behavior observed in the one-step RSB (1RSB) \cite{MezMon2009}. 
The peak value at $q=0$ and the peak position $q_*(t)$ are shown on the right side of Fig. \ref{fig:test3}. 
It appears that these approach finite values in the limit $t\rightarrow \infty$. 
In this work, we determine that RSB in the path ensemble occurs in the model under study.


The presence of two peaks in $P(q)$ indicates the coexistence of a high-overlap phase and a low-overlap phase in space-time.
When we introduce the generating function of $q$ as $\psi(\epsilon)\equiv \lim_{\tau\rightarrow \infty}\log\left\langle e^{\tau\epsilon q} \right\rangle/\tau$, which is the dynamical free energy of overlap, the coexistence leads to a singularity in the dynamical free energy at $\epsilon=0$, which implies a dynamical phase transition.
We note that such a dynamical phase transition can be described in terms of the path ensemble of two independent systems biased by overlap.
RSB in the trajectories indicates that the trajectories of two independent systems with a positive (zero) overlap are most weighted by an infinitely small positive (negative) bias $\epsilon$, which results in a singularity at $\epsilon=0$.


{\em Analysis.---}
We present arguments to support the occurrence of RSB in the path ensemble.
A key observation is that the ensemble of trajectories of the particle under a modified boundary condition is equivalent to the canonical ensemble of directed polymers with one end fixed that are subjected to a random potential. 
Explicitly, when we consider (\ref{eq:passive}) and (\ref{eq:KPZ}) with boundary conditions $x(\tau)=x_0=L/2$ and $\phi(x,0)=0$, the path probability density of the particle is written as 
\begin{eqnarray}
 \mathcal{P}\left[x | x(\tau)=x_0 \right] 
&=& \frac{1}{Z_\mathrm{DP}} 
e^{-\frac{1}{4D}\int_0^\tau dt \left[ \dot{x}(t)^2 -2 v\left( x(t), t \right) \right]},
\label{eq:fpp_dp}
\end{eqnarray}
where  $Z_\mathrm{DP}$ represents the normalization constant. 
The proof is as follows. 
We first write the Onsager--Machlup expression of the path probability density as
\begin{eqnarray}
\mathcal{P}\left[x | x(\tau)=x_0 \right] 
&=& \frac{1}{Z_0} 
e^{-\frac{1}{4D}\int_0^\tau dt 
\left[ \dot{x}(t) + 
\frac{\partial \phi}{\partial x}\left( x(t), t \right) \right]^2} 
\nonumber \\
&& \qquad \times e^{ - \frac{1}{2} 
\int_0^\tau dt \frac{\partial^2 \phi}{\partial x^2}\left( x(t), t \right)},
\label{eq:fpp_kpz}
\end{eqnarray}
where $Z_0$ denotes the normalization constant, and the second term in the exponential arises from the transformation from $\left[ \xi(t) \right]_{t=0}^\tau$ to $\left[ x(t) \right]_{t=0}^\tau$ with $x(\tau)$ fixed \cite{SM-Jacobian}. 
We subsequently rewrite (\ref{eq:fpp_kpz}) using (\ref{eq:KPZ}) as 
\begin{eqnarray}
&& \mathcal{P}\left[x | x(\tau)=x_0 \right] \nonumber \\
&=& \frac{1}{Z_0} 
e^{ -\frac{1}{4D}
\int_0^\tau dt \left[ 
\dot{x}(t)^2 + 2\frac{d}{dt}\phi\left( x(t),t \right) 
-2\frac{\partial \phi}{\partial t}
\left( x(t),t \right)\right]} \nonumber \\
&& 
\qquad \times  
e^{ -\frac{1}{4D}
\int_0^\tau dt \left\{\left[ \frac{\partial \phi}{\partial x}
\left( x(t), t \right) \right]^2 
+2 D \frac{\partial^2 \phi}{\partial x^2}
\left( x(t), t \right) \right\}}, \nonumber \\
&=& \frac{1}{Z_0} e^{-\frac{1}{2D}
\left[ 
\phi\left( x(\tau),\tau \right) 
- \phi\left( x(0),0 \right) 
\right] 
-\frac{1}{4D}
\int_0^\tau dt \left[ \dot{x}(t)^2 
-2 v\left( x(t), t \right) \right]}, \nonumber \\
&=& \frac{1}{Z_0} 
e^{-\frac{1}{2D}\phi\left( x_0,\tau \right)
 -\frac{1}{4D}
\int_0^\tau dt \left[ \dot{x}(t)^2 -2 v\left( x(t), t \right) \right]}.
 \label{eq:fpp_cal}
\end{eqnarray}
Since the first term of the exponential can be absorbed into the normalization constant, we obtain (\ref{eq:fpp_dp}).


The statistical model (\ref{eq:fpp_dp}) has been extensively studied \cite{KarZha1987, Kar1987, Par1990, Mez1990, BouOrl1990, HwaFis1994, Yos1996, BruDer2000, CLR2010, HalZha1995}. 
It has been shown that the system is in the frozen phase for $B>0$ and $D<\infty$.
The most important result here is that the replica symmetry of the system is broken \cite{Par1990, Mez1990}, which is obtained by the replica Bethe ansatz calculation and numerical calculation of the transfer matrix for a discretized model. 
Thus, the equivalence demonstrated above implies that the Langevin model with $x(\tau)$ fixed also exhibits RSB. 

\begin{figure}[t]
\includegraphics[clip, width=4.25cm]{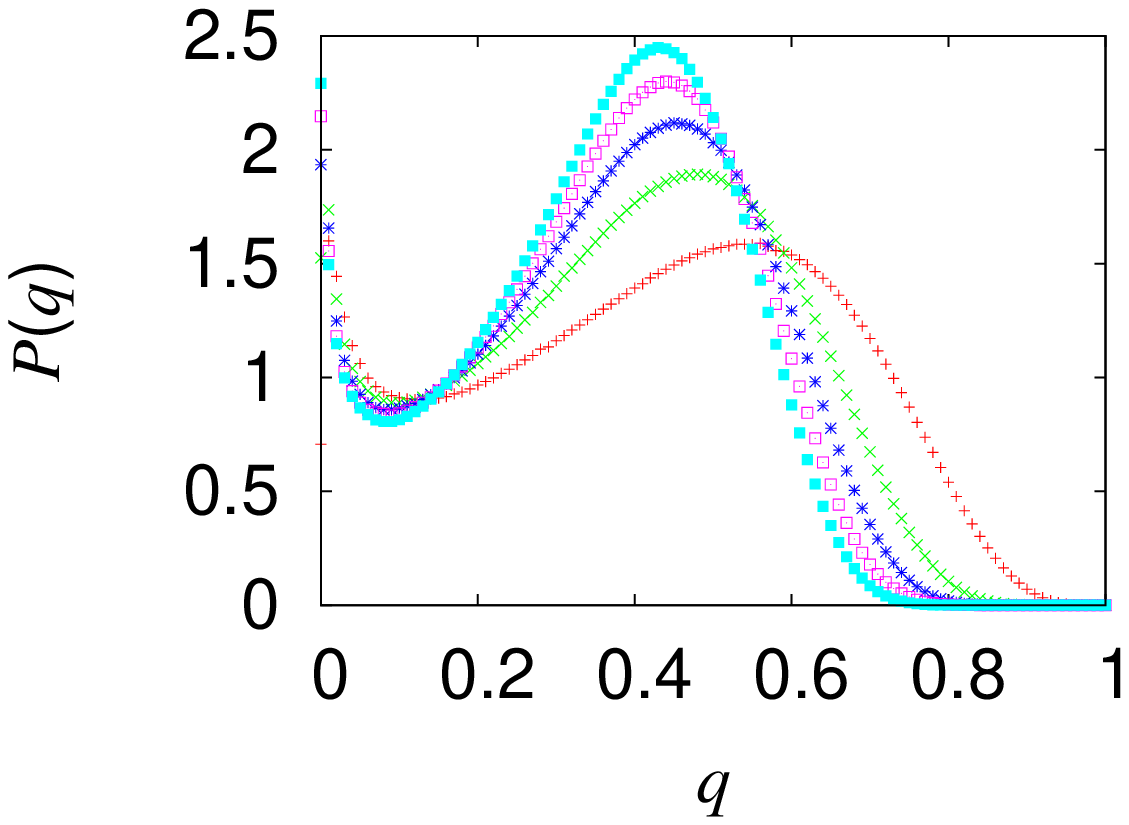}
\includegraphics[clip, width=4.25cm]{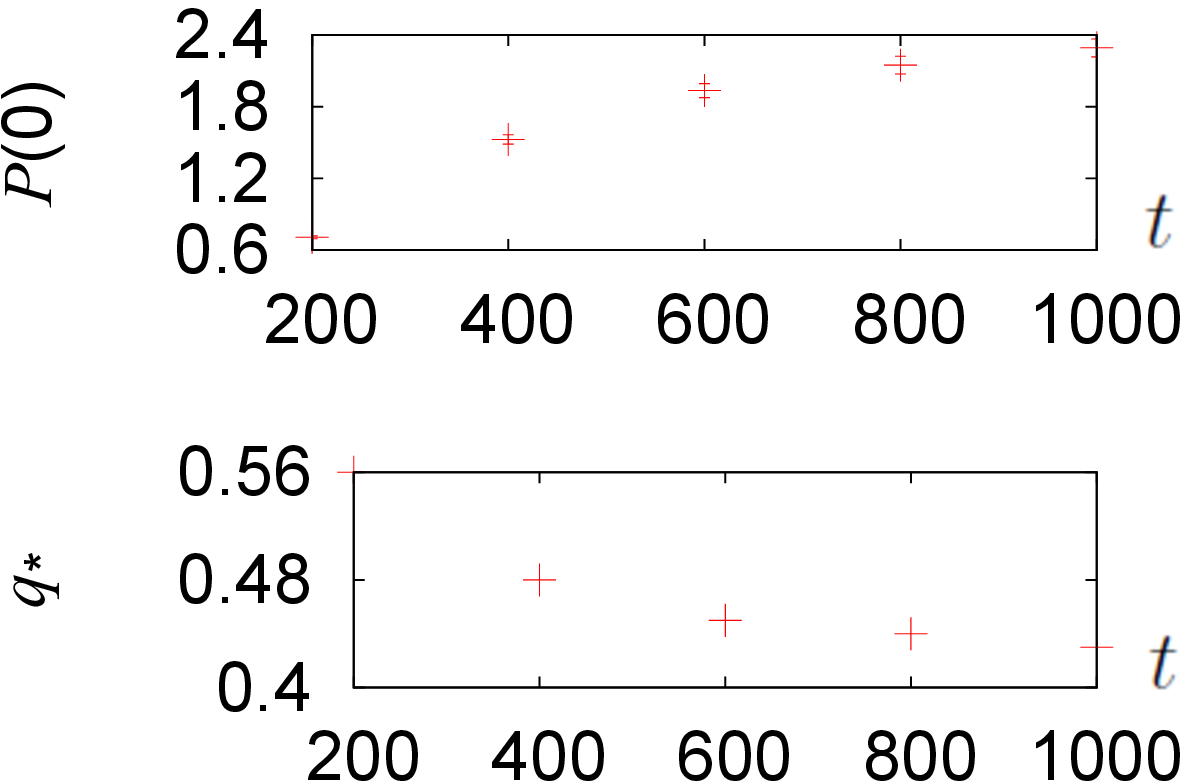}
\caption{(color online). Statistical properties of the ensemble (\ref{eq:fpp_kpz}). 
The distribution of the overlap, $P(q)$, for various $t^\prime\equiv \tau-t$ are plotted in the left side, and their peak values at $q=0$ and peak positions $q_*(t')$ are displayed in the right side.
The symbols and color labels in the graph with $t^\prime$ are the same as those in the graph with $t$ in Fig. \ref{fig:test3}}
\label{fig:test4}
\end{figure}

Therefore, if the particle behavior in the time region $1 \ll t\ll \tau$ is independent of $x(0)$ and $x(\tau)$, the original model with $x(0)$ fixed (which is easily prepared in experiments) can be shown to exhibit RSB. 
Although the validity of the independence of $x(0)$ and $x(\tau)$ is not assured, we can verify it in numerical simulations. 
That is, we numerically investigate the modified system with $x(\tau)$ fixed and compare the results with those obtained for the original model \cite{SM-numerical}.
On the left side of Fig. \ref{fig:test4}, we show $P(q)$ for the modified system.
This result is nearly identical to that of the original case. 
The peak value and the non-trivial peak position also behave in the same way as in the original model, as seen on the right side of Fig. \ref{fig:test4}. 
Thus, we can confidently conclude that $P(q)$ is independent of $x(0)$ and $x(\tau)$, and this implies that the original model definitely exhibits RSB.
See Ref. \cite{SM-weak} for discussions on the type of RSB.

At the end of our {\it Analysis}, we attempt to provide a physical picture of RSB in the path ensemble.
In Ref. \cite{Chi2002}, it was found for the $D=0$ case that valleys (local minima) of the velocity potential $\phi$ move continuously and eventually coalesce.
Therefore, when $D=0$, independent particles with slightly different initial conditions cluster in a specific valley, which leads to $P(q)$ concentrated on $q=1$.
However, in our model with $D>0$, a particle can escape from one valley of $\phi$ to another valley due to thermal noise $\xi$.
We subsequently explain that this activation effect is too weak to remove the concentration of $P(q)$ on a finite overlap.
First, it has been known that the steady-state distribution of the KPZ equation is equivalent to the probability distribution of a random potential field in the continuous version of the Sinai model \cite{BouGeo1990}, where ultraslow diffusion $\left\langle x(t)^2 \right\rangle \sim (\log{t})^4$ is observed.
Now, let us suppose that we observe the distance of two independent particles with different thermal noises during a time interval $t$.
The distance would be $\mathcal{O}\left( (\log{t})^2 \right)$ for large $t$ if the potential were frozen.
On the other hand, particles whose distance from each other is within a length scale $l_\mathrm{co}$ cluster in one valley by coalescences of valleys, where $l_\mathrm{co}$ is expected to be $\mathcal{O}\left( t^\alpha \right)$ with some $\alpha>0$ for large $t$ because it is determined by the diffusive nature of motion of the valleys.
Therefore, from $t^\alpha\gg (\log{t})^2$ for large $t$, we conclude that the phenomenon observed in the system with $D=0$ is robust to thermal noise.
The importance of the coalescence of valleys can be understood from the absence of RSB for the case where particles are driven by a field obeying the Edwards--Wilkinson (EW) equation \cite{SM-EW}, because the steady-state probability distribution of the potential field for the EW equation is the same as that for the KPZ equation while the valleys of the potential do not coalesce.


{\em Concluding Remarks.---}
We discuss previous studies related to our results.
It has been known that two independent and identical dynamical systems exhibit mutual attraction of trajectories with slightly different initial conditions when they are subject to a common noise.
Examples include synchronization of two independent oscillators driven by a common noise \cite{BKOVZ2002, TerTan2004}, and aggregation of independent particles driven by a common Gaussian random velocity field \cite{Deu1985, WilMeh2003} or a common spatio-temporal chaotic field \cite{BohPik1993}.
A more recent work on diffusive fluctuating hydrodynamics reports a similar phenomenon through the analysis of the dynamical free energy of the Lyapunov exponent \cite{LSTW2015}.
These phenomena are similar to that studied in our paper in that attraction of trajectories in two independent and identical systems is concerned.
One can interpret that a common disordered velocity field in our study corresponds to a common noise in previous studies.
Although this interpretation is reasonable, the two systems in our study are subject to independent thermal noises in addition to a common velocity field, which makes a contrast to previous studies.
We also remark that mutual attraction of trajectories in two independent systems was observed in our system without thermal noise \cite{Chi2002}.
We emphasize that to find the common-noise-induced attraction robust against independent noises is a highly non-trivial problem, which corresponds to the discovery of a phase transition at finite temperature for a symmetry breaking at $T=0$ in equilibrium statistical mechanics.
The main achievement of our study is that we solve this problem by using the concept of RSB.


In sum, we have demonstrated RSB in trajectories for a tracer particle passively driven by a field obeying the noisy Burgers equation. 
In this model, two tracer particles driven by a common velocity field are always close to each other when they freeze to the same trajectory, while they become separated from each other with time when each particle freezes to a different trajectory.
These two cases correspond to two peaks in $P(q)$ at $q=q_*$ and $q=0$, respectively.
Such a pathological transportation property is detected only by the observation of $P(q)$.
We believe that the dynamical free energy of overlap will be a useful tool to characterize a singularity of trajectories for a wide class of phenomena including chemical networks \cite{Kau1969} and cell differentiation \cite{Sinet2014, Kan2006}.


The authors thank T. Nemoto for useful discussions, particularly regarding the derivation of (\ref{eq:fpp_dp}). 
They also thank F. van Wijland and \'{E}. Fodor for valuable comments.
The present study was supported by KAKENHI (Nos. 25103002 and 26610115), a Grant-in-Aid for JSPS Fellows (No. 2681), and by the JSPS Core-to-Core program ``Non-equilibrium dynamics of soft-matter and information.''


\clearpage
\widetext
\setcounter{equation}{0}
\def\theequation{S\arabic{equation}}
\setcounter{figure}{0}
\def\thefigure{S\arabic{figure}}
\appendix

\section{SUPPLEMENTAL MATERIAL}


\section{1. Estimation of $d_0$}

\begin{figure}[htbp]
\includegraphics[clip, width=8.0cm]{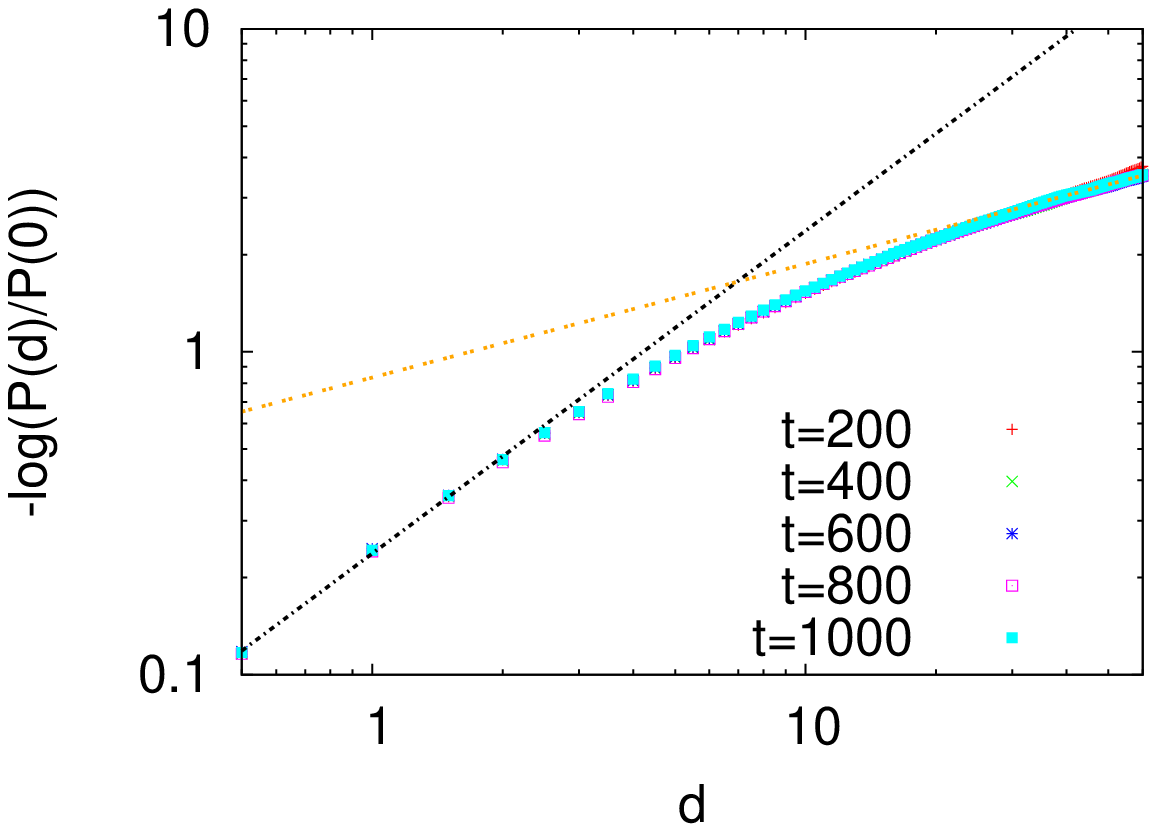}
\caption{(color online). A log--log plot of $-\log\left[P(d)/P(0)\right]$ in the range $d\in[0,60]$.
The two straight lines represent (\ref{eq:exp}) with $d_0=4.2$ and (\ref{eq:str_exp}) with $\alpha = 0.35$ and $b=1.7$, respectively.}
\label{fig:str_exp}
\end{figure}

We estimate the characteristic length $d_0$ of the time-independent region $(t >200)$  of $P(d)$ in Fig. \ref{fig:test2} in the main text. 
Note that a logarithmic scale was used for the vertical axis.
In Fig. \ref{fig:str_exp}, $\log[{P(d)}/{P(0)}]$ is plotted as a function of $d$.
We find two different behaviors: the exponential form in small $d$ $(d<2)$
\begin{eqnarray}
 \log\frac{P(d)}{P(0)} &=& - \frac{d}{d_0},
 \label{eq:exp}
\end{eqnarray}
and the stretched exponential form
\begin{eqnarray}
 \log\frac{P(d)}{P(0)} &=& - \left(\frac{d}{b}\right)^\alpha
 \label{eq:str_exp}
\end{eqnarray}
in large $d$ $(20<d<60)$.
We interpret that two particles are trapped when their distance is less than $d_0$.
We thus characterize the localization length by $d_0$.
Two straight lines in Fig. \ref{fig:str_exp} represent (\ref{eq:exp}) with $d_0=4.2$ and (\ref{eq:str_exp}) with $\alpha = 0.35$ and $b=1.7$, respectively.


\section{2. Derivation of (\ref{eq:fpp_kpz})}

We derive the Onsager--Machlup expression of path probability density with the final position $x(\tau)$ fixed.
We first discretize the Langevin equation (\ref{eq:passive}) as
\begin{eqnarray}
 x_{n+1}-x_{n} &=& u\left( \frac{x_{n+1}+x_{n}}{2}, n\Delta t \right) \Delta t + \sqrt{2D}\Delta W_{n},
 \label{eq:discretized_Langevin}
\end{eqnarray}
where $\Delta t \equiv \tau/M$ and $\Delta W_{n}$ is the Wiener increment.
It should be noted that $n$ takes the value $0\leq n\leq M-1$.
The path probability density with the final position fixed is then written as
\begin{eqnarray}
 \mathcal{P}\left( x_{0}, \cdots, x_{M-1} | x_{M} \right) &=& \prod_{n=0}^{M-1} \left| \frac{\partial \Delta W_n}{\partial x_n} \right| \left( \frac{1}{\sqrt{4\pi D \Delta t}} \right)^M e^{-\sum_{n=0}^{M-1}\frac{\left( \Delta W_n \right)^2}{4D\Delta t}},
\end{eqnarray}
where $\prod_{n=0}^{M-1} \left| {\partial \Delta W_n}/{\partial x_n} \right|$ is the Jacobian of the transformation from $\left( x_{0}, \cdots, x_{M-1} \right)$ to $\left( \Delta W_{0}, \cdots, \Delta W_{M-1} \right)$.
By using (\ref{eq:discretized_Langevin}), we calculate
\begin{eqnarray}
 \left| \frac{\partial \Delta W_{n}}{\partial x_{n}} \right| &=& \frac{1}{\sqrt{2D}}\left[ 1+\frac{\Delta t}{2}\frac{\partial u}{\partial x}\left( \frac{x_{n+1}+x_{n}}{2}, n\Delta t \right) \right] \nonumber \\
 &\simeq& \frac{1}{\sqrt{2D}}e^{\frac{\Delta t}{2}\frac{\partial u}{\partial x}\left( \frac{x_{n+1}+x_{n}}{2}, n\Delta t \right)}.
\end{eqnarray}
Taking the limit $\Delta t \rightarrow 0$, we obtain (\ref{eq:fpp_kpz}) in the main text.

We note that the Jacobian term in (\ref{eq:fpp_kpz}) is different from a standard one for the path probability density with the initial position $x(0)$ fixed.
The path probability density with the initial position fixed is written as
\begin{eqnarray}
 \mathcal{P}\left( x_{1}, \cdots, x_{M} | x_{0} \right) &=& \prod_{n=0}^{M-1} \left| \frac{\partial \Delta W_n}{\partial x_{n+1}} \right| \left( \frac{1}{\sqrt{4\pi D \Delta t}} \right)^M e^{-\sum_{n=0}^{M-1}\frac{\left( \Delta W_n \right)^2}{4D\Delta t}},
\end{eqnarray}
where $\prod_{n=0}^{M-1} \left| {\partial \Delta W_n}/{\partial x_{n+1}} \right|$ is the Jacobian of the transformation from $\left( x_{1}, \cdots, x_{M} \right)$ to $\left( \Delta W_{0}, \cdots, \Delta W_{M-1} \right)$.
By using (\ref{eq:discretized_Langevin}), we calculate
\begin{eqnarray}
 \left| \frac{\partial \Delta W_{n}}{\partial x_{n+1}} \right| &=& \frac{1}{\sqrt{2D}}\left[ 1-\frac{\Delta t}{2}\frac{\partial u}{\partial x}\left( \frac{x_{n+1}+x_{n}}{2}, n\Delta t \right) \right] \nonumber \\
 &\simeq& \frac{1}{\sqrt{2D}}e^{-\frac{\Delta t}{2}\frac{\partial u}{\partial x}\left( \frac{x_{n+1}+x_{n}}{2}, n\Delta t \right)}.
\end{eqnarray}
Taking the limit $\Delta t \rightarrow 0$, we obtain the standard Onsager--Machlup expression.


\section{3. Numerical realization of (\ref{eq:fpp_kpz})}

We explain a numerical method for obtaining the path ensemble given by (\ref{eq:fpp_kpz}) where the final condition $x(\tau)$ is fixed.
First, we calculate $\phi(x,t)$, which obeys the KPZ equation (\ref{eq:KPZ}) with the initial condition $\phi(x,0)=0$, during a time interval $t\in[0,\tau]$ and store the values of $\phi(x,t)$ for all $x$ and $t$.
Then we calculate $\tilde{x}(t)\equiv x(\tau-t)$ from the Langevin equation 
\begin{eqnarray}
 \dot{\tilde{x}}(t) &=& 
\frac{\partial \phi}{\partial x}\left( \tilde{x}(t), \tau-t \right) + \xi(t)
\end{eqnarray}
with the fixed initial condition $\tilde{x}(0)=x_0$.
The path probability density of $\tilde{x}(t)$ is given by
\begin{eqnarray}
\tilde{\mathcal{P}}\left[\tilde{x} | \tilde{x}(0)
=x_0 \right] &=& \frac{1}{Z_0} 
e^{-\frac{1}{4D}\int_0^\tau dt 
\left[ \dot{\tilde{x}}(t) - 
\frac{\partial \phi}{\partial x}
\left( \tilde{x}(t), \tau-t \right) \right]^2 - 
\frac{1}{2} \int_0^\tau dt \frac{\partial^2 \phi}{\partial x^2}
\left( \tilde{x}(t), \tau-t \right)},
\end{eqnarray}
which is indeed equal to (\ref{eq:fpp_kpz}) due to the definition $\tilde{x}(t)= x(\tau-t)$.
In this way, we can construct the path ensemble (\ref{eq:fpp_kpz}).


\section{4. Type of RSB} 


We briefly review classification of RSB. 
In general, RSB in a system is detected by the nontriviality of $P(q)$. 
The graph of $P(q)$ in a RSB phase takes one of several forms \cite{MezMon2009sm}. 
For the full RSB case, $P(q)$ is a broad function even in the limit $\tau\rightarrow \infty$. 
For the one-step RSB (1RSB) case, $P(q)$ converges to a sum of two $\delta$-functions in the limit $\tau\rightarrow \infty$.
In addition to these well-known cases, there is a case that $\lim_{\tau\rightarrow \infty}P(q)$ has a single peak, but the replica symmetry of the system is broken in the sense that there are many metastable states whose free energy per degree of freedom is equal to that of the lowest free energy state \cite{HusFis1987sm, ParVir1989sm}. 
This case is called a weak breaking of the replica symmetry \cite{Par1990sm, Mez1990sm}.

\begin{figure}[htbp]
\includegraphics[clip, width=8.0cm]{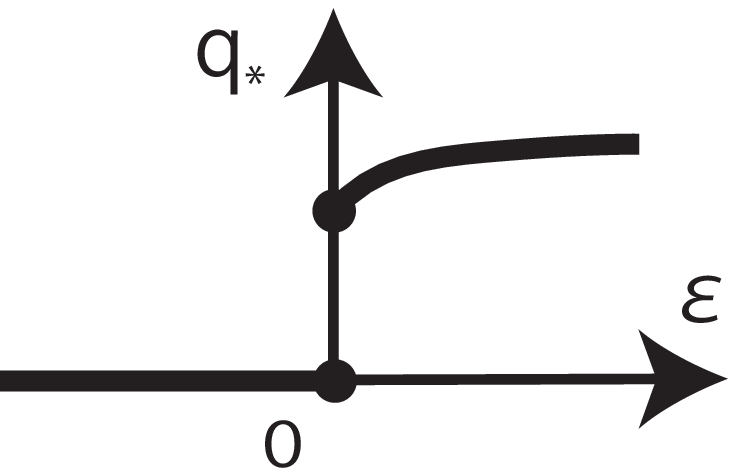}
\includegraphics[clip, width=8.0cm]{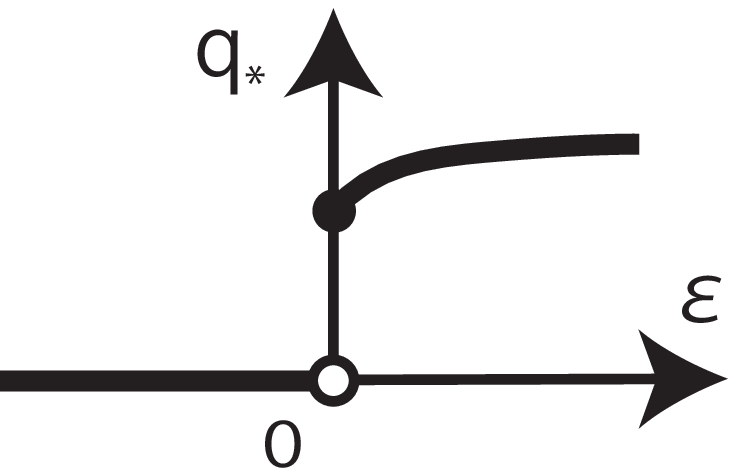}
\caption{Schematic graphs of the peak position $q_*$ of $P(q)$ for the 1RSB case (left) and the weak breaking case (right).}
\label{fig:epsilon-q}
\end{figure}

More explicitly, we can formulate and classify RSB by considering an $\epsilon$-biased ensemble with the biasing factor $e^{\tau \epsilon q}$.
That is, by defining the scaled cumulant generating function 
\begin{eqnarray}
\psi(\epsilon) \equiv 
\lim_{\tau\rightarrow \infty} \frac{1}{\tau}
\log \left\langle e^{\tau \epsilon q} \right\rangle, 
\end{eqnarray}
we consider the expectation of the overlap in the $\epsilon$-biased ensemble
\begin{eqnarray}
\left\langle q \right\rangle_\epsilon 
\equiv \frac{\partial }{\partial \epsilon}\psi(\epsilon).
\end{eqnarray}
We identify RSB if $\lim_{\epsilon\rightarrow +0} \left\langle q \right\rangle_{\epsilon} \neq \lim_{\epsilon\rightarrow -0} \left\langle q \right\rangle_{\epsilon}$. 
In particular, the overlap $q$ in the 1RSB case, takes two values at $\epsilon=0$ depending on the limit $\epsilon \to +0$ or $\epsilon \to -0$, while the overlap in the weak breaking case takes only one of them in the limit $\tau \to \infty$. 
See Fig. \ref{fig:epsilon-q} as schematic graphs representing this fact.


Here, we remark on the type of RSB of directed polymers.
The numerical result for $P(q)$ in the main text suggests 1RSB, because it contains two peaks at $q=0$ and $q = q_{\ast} \neq 0$ that become sharper with time. 
However, it has been previously pointed out that the replica symmetry of the model (\ref{eq:fpp_dp}) is weakly broken \cite{Par1990sm, Mez1990sm}, which is in contrast to 1RSB or full RSB that is familiar in the theory of mean-field spin glass models.  
More precisely, while the free energy of the system is calculated from the replica symmetric ansatz, the expectation of the overlap between two identical copies coupled together with strength $\epsilon$ experiences a jump at $\epsilon=0$. 
Furthermore, it has been conjectured in Ref. \cite{Par1990sm} that $P(q)=\delta(q-q_*)$ with $q_*\neq 0$ in the limit $\tau\rightarrow \infty$, while the second peak of $P(q)$ at $q=0$ for finite $\tau$ exhibits slow relaxation of the $\tau^{-1/3}$ type.
This means that the path ensemble is dominated by a single trajectory in $\tau\rightarrow \infty$ but there are many metastable states whose free energy per degree of freedom is equal to that of the lowest free energy state.
See Ref. \cite{Mez1990sm} for a related argument.
We also confirmed such slowly decaying behavior of the peak at $q=0$ for a discretized version of directed polymers in the next section.
This is qualitatively different from the behaviors depicted in Figs. \ref{fig:test3} and \ref{fig:test4}.
Therefore, to the extent of our investigation, we cannot exclude the possibility of 1RSB for both the original model and the statistical model (\ref{eq:fpp_kpz}).
Further numerical analysis will be required to resolve this problem.


\section{5. Discretized version of directed polymers}


We provide numerical results for $P(q)$ for a discretized version of directed polymers.
We consider a polymer on the square lattice as a path $[y]\equiv(y_0,y_1,\cdots,y_M)$ with $y_j\in \{1, \cdots, N\}$ satisfying the constraint $y_{j+1}\in\left\{ y_j-1,y_j,y_j+1 \right\}$.
The periodic boundary conditions are used in the $y$ direction.
The polymer is subjected to a zero mean Gaussian random potential $v(y,j)$ with variance $\left\langle v(y,j) v(y^\prime,j^\prime) \right\rangle = A\delta_{y,y^\prime}\delta_{j,j^\prime}$.
We assume the following equilibrium distribution of the polymer with one end fixed:
\begin{eqnarray}
 \mathcal{P}_\mathrm{DP}\left[y|y_0=a\right] 
&=& \frac{1}{Z} \prod_{j=0}^{M-1} e^{- v(y_{j+1},j+1)} 
\left[ \delta_{y_{j+1},y_{j}} + \gamma \delta_{y_{j+1},y_{j}+1} 
+ \gamma \delta_{y_{j+1},y_{j}-1} \right],
\label{eq:pp_ddp}
\end{eqnarray}
where $Z$ is the normalization constant and $\gamma$ is a parameter related to an elastic constant \cite{Mez1990sm}.
The partition function $Z$ of this model was calculated by using a method of the transfer matrix \cite{HalZha1995sm}
\begin{eqnarray}
Z\left( y, j+1 \right) &=& \sum_{y^\prime} 
T_{j+1}\left(y|y^\prime \right) Z\left( y^\prime, j \right), \\
 T_{j+1}\left(y|y^\prime \right) &\equiv& 
\frac{1}{1+2\gamma}e^{- v(y,j+1)} 
\left[ \delta_{y,y^\prime} + \gamma \delta_{y,y^\prime+1} 
+ \gamma \delta_{y,y^\prime-1} \right]
\end{eqnarray}
with the initial condition $Z\left( y, 0 \right)=\delta_{y,a}$.

\begin{figure}[htbp]
\includegraphics[clip, width=8.0cm]{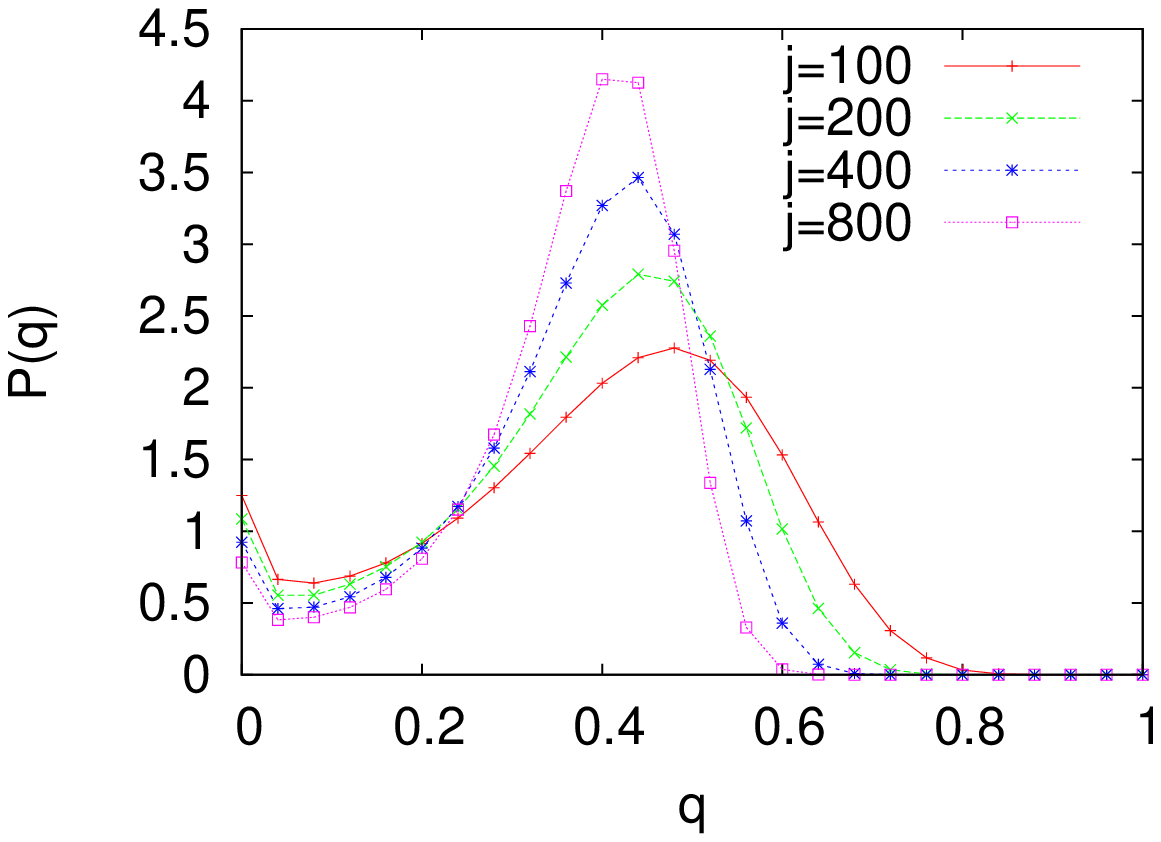}
\includegraphics[clip, width=8.0cm]{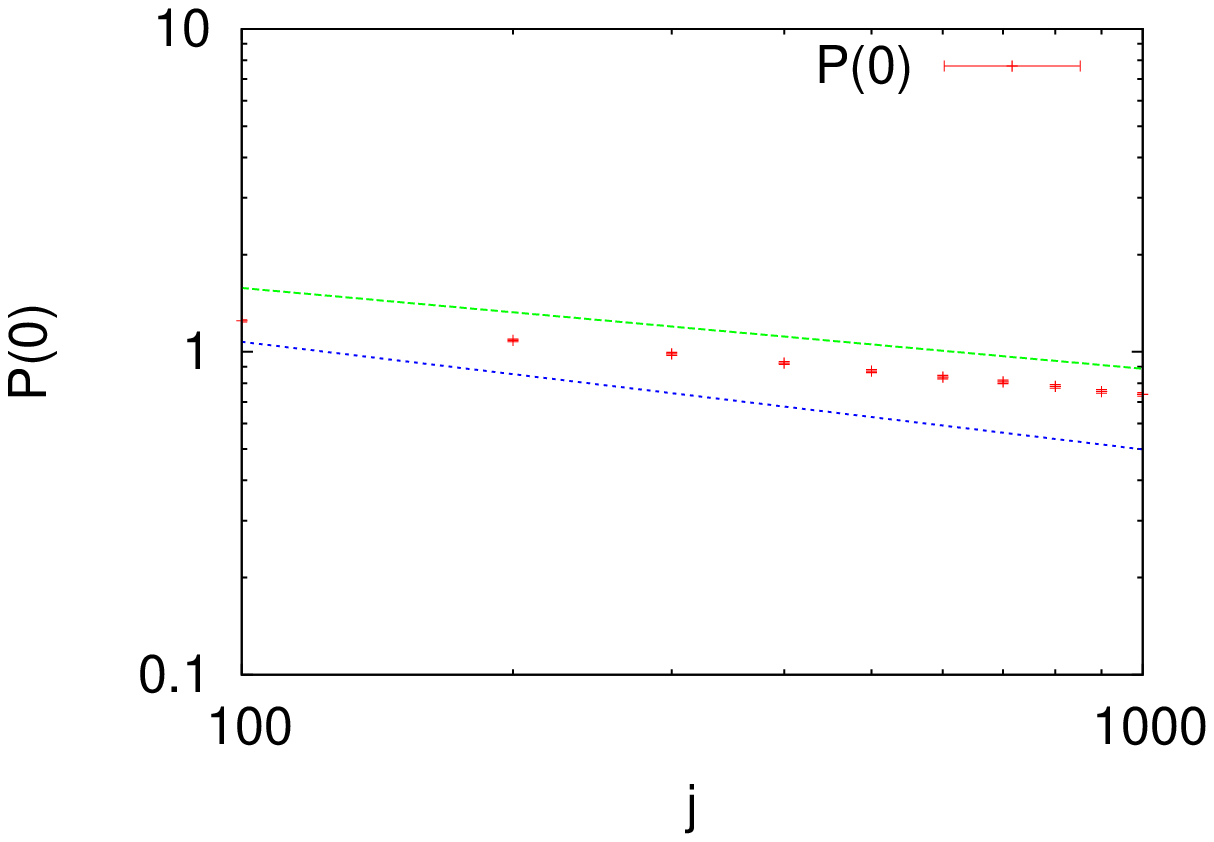}
\caption{(color online). The left figure shows the overlap distribution $P(q)$ for the discretized model. The right figure shows the time dependence of $P(q=0)$ in a log--log plot. The two straight lines represent $P(0)\sim j^{-1/4}$ and $P(0)\sim j^{-1/3}$, respectively.}
\label{fig:discrete}
\end{figure}

We define the overlap between two trajectories $\left[y^{(1)}\right]$ and $\left[y^{(2)}\right]$ as
\begin{eqnarray}
q\left( \left[y^{(1)}\right], \left[y^{(2)}\right] \right) 
&\equiv& \frac{1}{M} \sum_{j=1}^M \delta_{y^{(1)}_j, y^{(2)}_j}.
\end{eqnarray}
The distribution of the overlap  $P(q)$ for this model was measured numerically. 
The parameter values used in the calculation are $A=1.0$ and $\gamma=0.1$.
The statistical average was taken over $80000$ samples for one fixed realization of the potential, and the disorder average with respect to random potentials was calculated from $2000$ samples. 
The result is shown on the left side of Fig. \ref{fig:discrete}. 
Two peaks are observed similarly to those in the main text, but the important difference is that, as shown on the right side of Fig. \ref{fig:discrete}, the peak at $q=0$ decreases as a power law function $\tau^{-\mu}$, which is consistent with the prediction in Refs. \cite{Par1990sm, Mez1990sm}. 
More precisely, although the value of $\mu$ estimated from the numerical data is closer to $1/4$ than $1/3$ proposed in previous studies \cite{Par1990sm, Mez1990sm}, we conjecture that this slight discrepancy comes from finite size effects. 
It should be noted that the increasing behavior of $P(q=0)$ on the right side of Fig. \ref{fig:test4} in the main text has never been observed in this model. 


For completeness, we explain our calculation method of $P(q)$, which relies on a Markov chain model whose path probability is equivalent to (\ref{eq:pp_ddp}). 
Below, we shall explain the Markov chain explicitly.


We construct a Markov chain by extending a method proposed in Refs. \cite{JacSol2010sm, NemSas2011sm} to that for time-dependent cases \cite{Nemunpsm}.
First, we define a function $\Phi(y,j)$ by an equation
\begin{eqnarray}
\Phi\left( y^\prime,j \right) 
&=& \sum_y \Phi\left( y,j+1 \right) T_{j+1}\left(y|y^\prime \right)
 \label{eq:discrete_dynamics_field}
\end{eqnarray}
with the final condition $\Phi\left( y,M \right) = 1$.
Next, by defining 
\begin{eqnarray}
\tilde{T}_{j+1}\left(y|y^\prime \right) 
&\equiv& \frac{\Phi\left( y,j+1 \right)}
{\Phi\left( y^\prime,j \right)} T_{j+1}\left(y|y^\prime \right),
\end{eqnarray}
we can confirm $\sum_y\tilde{T}_{j+1}\left(y|y^\prime \right)=1$ from the definition of $\Phi$. 
Thus, by identifying $\tilde T$ as a time dependent transition probability, we have a Markov chain
\begin{eqnarray}
P\left( y,j+1 \right) 
&=& \sum_{y^\prime} \tilde{T}_{j+1}
\left(y|y^\prime \right) P\left( y^\prime,j \right).
\label{eq:discrete_dynamics}
\end{eqnarray}
We fix the initial condition as $P(y,0)=\delta_{y,a}$ with some $a$.


Now, we write a realization of state at time $j$ as $y_j$ and denote a path by $[y]\equiv(y_0,y_1,\cdots,y_M)$.
The probability of path is given by
\begin{eqnarray}
\mathcal{P}\left[y|y_0=a\right] 
&\equiv& \prod_{j=0}^{M-1} \tilde{T}_{j+1}\left(y_{j+1}|y_j \right).
 \label{eq:bpp_dkpz}
\end{eqnarray}
This can be rewritten as
\begin{eqnarray}
 \mathcal{P}\left[y|y_0=a\right] 
&=& \frac{\Phi\left( y_M,M \right)}
{\Phi\left( y_0,0 \right)}\prod_{j=0}^{M-1} 
T_{j+1}\left(y_{j+1}|y_j \right) \nonumber \\
&=& \frac{1}{\Phi\left( a,0 \right)}
\prod_{j=0}^{M-1} T_{j+1}\left(y_{j+1}|y_j \right) \nonumber \\
&=& \frac{1}{Z} \prod_{j=0}^{M-1} e^{- v(y_{j+1},j+1)} 
\left[ \delta_{y_{j+1},y_{j}} + \gamma \delta_{y_{j+1},y_{j}+1} 
+ \gamma \delta_{y_{j+1},y_{j}-1} \right],
\end{eqnarray}
where $Z$ is the normalization constant.
This path probability is equivalent to (\ref{eq:pp_ddp}).
In numerical calculation, $\Phi\left( y,j \right)$ is calculated by (\ref{eq:discrete_dynamics_field}), and then $\mathcal{P}\left[y|y_0=a\right]$ is obtained by (\ref{eq:bpp_dkpz}).
By collecting trajectories numerically, we obtained $P(q)$.


\section{6. Results for the Edwards--Wilkinson equation}

\begin{figure}[htbp]
\includegraphics[clip, width=8.0cm]{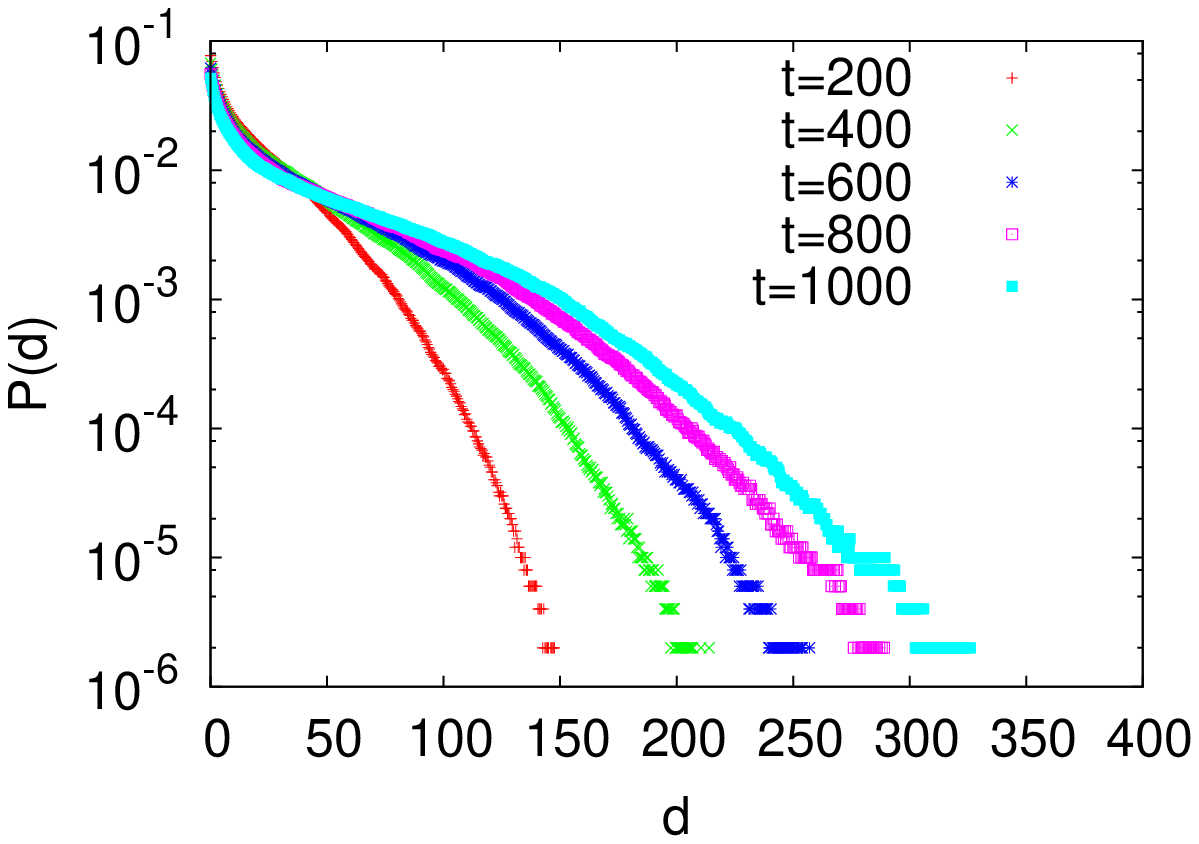}
\includegraphics[clip, width=8.0cm]{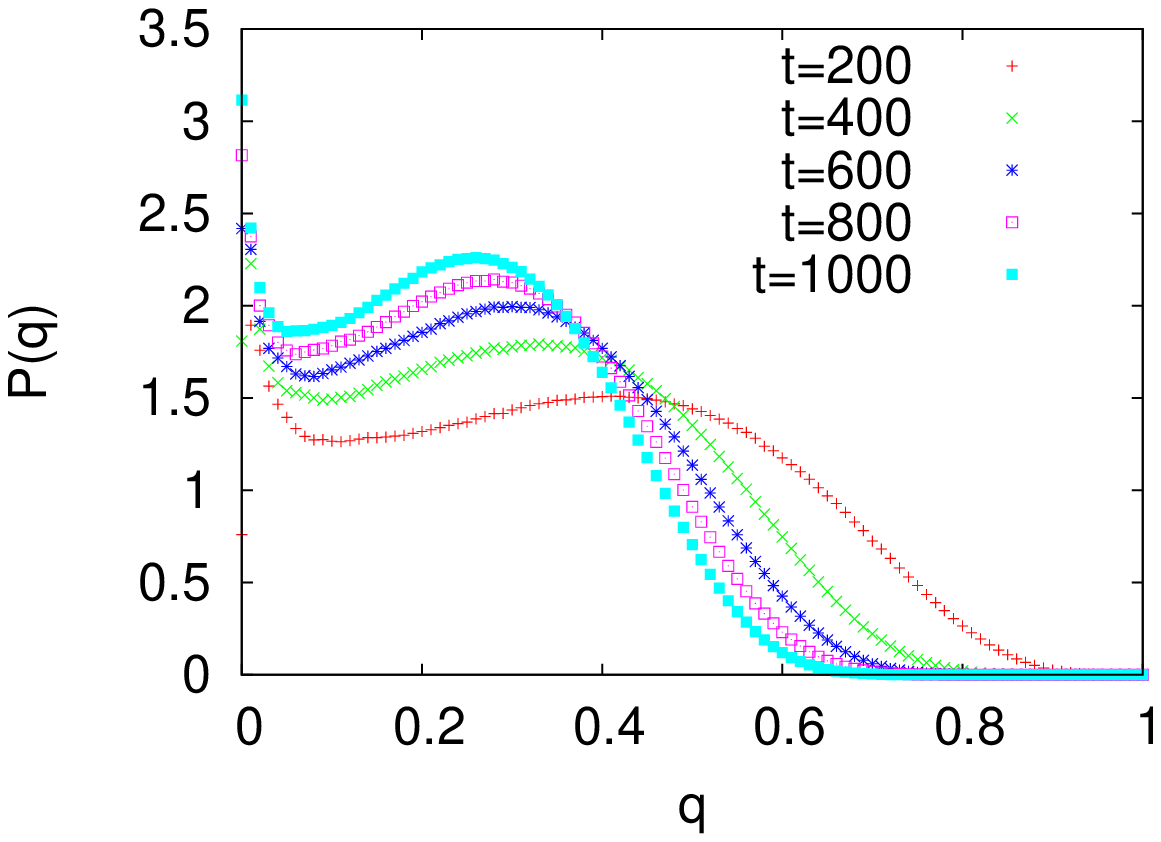}
\caption{(color online). Statistical quantities for a Brownian particle driven by a field obeying the EW equation. 
The distribution of the relative distance, $P(d)$ (left), and the distribution of the overlap, $P(q)$ (right).}
\label{fig:EW}
\end{figure}

As a control experiment, we study a passive particle driven by a field obeying the Edwards--Wilkinson (EW) equation \cite{EW1982sm}
\begin{eqnarray}
\frac{\partial \phi}{\partial t}\left( x,t \right) 
&=& D\frac{\partial^2 \phi}{\partial x^2}\left( x,t \right) + v(x,t).
\end{eqnarray}
The distribution of the relative distance, $P(d)$, displayed on the left side of Fig. \ref{fig:EW} exhibits non-Gaussian behavior, which is similar to the right side of Fig. \ref{fig:test2} in the main text.
However, the shape of the graph $P(q)$ displayed on the right side of Fig. \ref{fig:EW} is qualitatively different from that on the left side of Fig. \ref{fig:test3}, because the local minimum between the two $P(q)$ peaks of Fig. \ref{fig:EW} increases with time.
Since the non-trivial peak would be absorbed into the peak at $q=0$ in the limit $\tau \rightarrow \infty$, we expect that there is no RSB for the EW case. 


\begin{thebibliography}{99}
\bibitem{Nemunpsm} The construction method of such Markov processes for time-dependent cases was provided by T. Nemoto.
\end{thebibliography}
\end{document}